\documentclass[aps,superscriptaddress,twocolumn]{revtex4-2}
\usepackage{amsmath}    
\usepackage{graphicx}   
\usepackage{verbatim}   
\usepackage{color}      
\usepackage{subfigure}  
\usepackage{hyperref}   
\usepackage{bm}
\usepackage{epstopdf}
\graphicspath{{./figures/}}
\usepackage{float}
\usepackage{chngcntr} 
\usepackage{xcolor}

\usepackage{bibentry}

\newcommand{\ignore}[1]{}
\newcommand{\nobibentry}[1]{{\let\nocite\ignore\bibentry{#1}}}
\newcommand{\bibfnamefont}[1]{#1}
\newcommand{\bibnamefont}[1]{#1}

\newcommand{\f}[1]{\textit{#1}}
\newcommand{\bi}[1]{\textbf{\textit{#1}}}
\newcommand{\bo}[1]{\textbf{#1}}

\begin{document}

\title{Spin current generation driven by skyrmion dynamics\\ under magnetic anisotropy and polarized microwaves}

\author{Seno Aji}
\email{senji77@sci.ui.ac.id}
\affiliation{Department of Physics, Faculty of Mathematics and Natural Sciences, Universitas Indonesia, Depok 16424, Indonesia.}
\author{Muhammad Anin Nabail Azhiim}
\affiliation{Department of Physics, Faculty of Mathematics and Natural Sciences, Universitas Indonesia, Depok 16424, Indonesia.}
\author{Nur Ika Puji Ayu}
\affiliation{Department of Physics, Faculty of Mathematics and Natural Sciences, Universitas Indonesia, Depok 16424, Indonesia.}
\author{Adam Badra Cahaya}
\affiliation{Department of Physics, Faculty of Mathematics and Natural Sciences, Universitas Indonesia, Depok 16424, Indonesia.}
\author{Koichi Kusakabe}
\affiliation{School of Science, Graduate School of Science, University of Hyogo, 3-2-1 Kouto, Kamigori-cho, Ako-gun, Hyogo, 678-1297, Japan.}
\author{Muhammad Aziz Majidi}
\email{aziz.majidi@sci.ui.ac.id}
\affiliation{Department of Physics, Faculty of Mathematics and Natural Sciences, Universitas Indonesia, Depok 16424, Indonesia.}


\begin{abstract}
We have investigated the spin-current pumped by the skyrmion-host material with the lack of inversion symmetry through the microwave resonance process. The effects of magnetic anisotropy and polarized microwaves are examined by micromagnetic simulations. Our results reveal two distinct skyrmion phases, designated as SkX type-I and II, which emerge at low~($K_z~<~0.1$~meV) and high ($K_z>0.1$~meV) magnetic anisotropy constants, respectively, having different characteristics of spin excitations. The SkX type-I exhibits spin dynamics where the resonant frequency of the breathing mode is lying in between the clockwise and counterclockwise gyration modes of Bloch-type skyrmion at a very low anisotropy, and is crossing over the counterclockwise mode at $K_z \sim 0.04$~meV. Meanwhile, the SkX type-II exhibits distinct spin excitations in which the clockwise mode is notably absent, while the counterclockwise modes exist at both low and high resonant frequencies. This suggests that the magnetic anisotropy plays an essential role in the spin dynamics. Furthermore, the resulting spin excitations induce spin currents with exotic features under the polarized microwaves. The spin currents induced, for instance, by low-lying in-plane excitations are strongly enhanced under the left-handed circularly polarized microwaves, but quenched by the right-handed circularly polarized microwaves regardless of the sign of the Dzyaloshinskii-Moriya interaction. These results may pave the way for understanding the non-trivial interplay between magnetic anisotropy and polarized microwaves in the generation of spin currents by a resonant process. 
   

\end{abstract}
\maketitle

\section{Introduction}

Magnetic skyrmions are vortex-like spin configurations with nontrivial topology and particle-like properties in continuous field theory \cite{SKYRME1962556}. They crystallize in the skyrmion lattice with typical lattice constants from 10 to 100 nm, and give rise to emergent electrodynamics such as the topological Hall effect owing to a finite scalar spin chirality \cite{PhysRevLett.102.186602,PhysRevLett.106.156603,Schulz2012,Nagaosa2013,Ritz2013,Chen2017,Jiang2017}. Due to their exceptional stability against continuous deformation, they are potentially applicable for the spintronic devices with high information density ~\cite{Fert2013,Rosch2013}. The first observation of magnetic skyrmions was found in MnSi by the small-angle neutron scattering (SANS) experiments in 2009. This B20-type cubic noncentrosymmetric chiral magnet has the skyrmion lattice with a hexagonal pattern observed in the reciprocal space~\cite{muhlbauer}. A few years later, other B20 materials were also reported having skyrmion lattice phase, such as in metallic MnGe, FeGe, semiconductor Fe$_{0.8}$Co$_{0.2}$Si, and insulator Cu$_2$OSeO$_3$~\cite{Nagaosa2013}. Recent reports indicate that the magnetic skyrmions were also observed experimentally in the centrosymmetric magnet materials \cite{kurumaji,Hirschberger2019,Takagi2022}.

\begin{figure}[b]
\centering
\includegraphics[width=6.5cm]{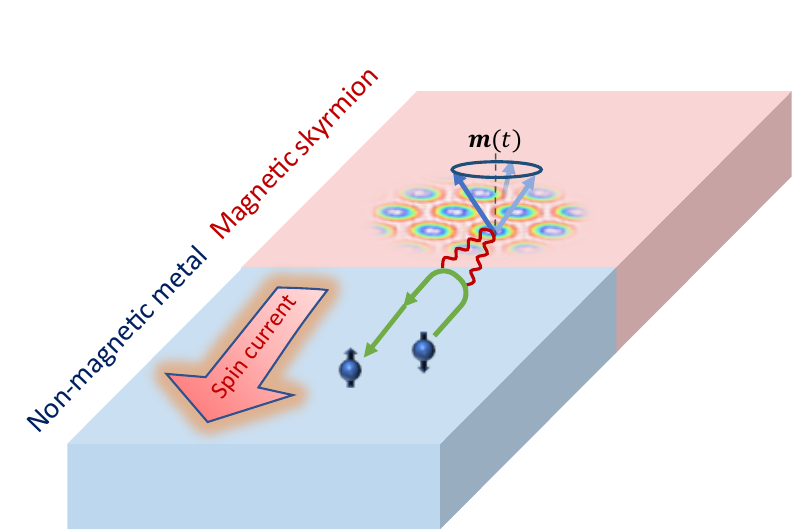}
\caption{Illustration of spin-current pumped by the collective excitations of the magnetic skyrmion. The generated spin-current flows through non-magnetic material.}
\label{fig:ilus_sp}
\end{figure}

An investigation of magnetic skyrmions has gained more interest since the theoretical prediction discovers that the magnetic skyrmions can be collectively excited or resonated with microwaves~\cite{mochizuki}. Such collective excitations were observed experimentally in the insulating chiral magnet Cu$_2$OSeO$_3$~\cite{onose_lett,Okamura2013,PhysRevLett.114.197202}. Later, Bloch-type skyrmion lattices in various metallic, semiconducting, and insulating chiral magnets were reported to have commonly three types of collective spin excitation characterized as clockwise (CW), counterclockwise (CCW), and breathing modes~\cite{Schwarze2015}. Moreover, it was reported that the microwave resonance process of the magnetic skyrmion can induce spin currents by using spin-pumping method \cite{hirobe,Zhang_2020}. This method was firstly introduced in thin ferromagnetic films~\cite{PhysRevLett.88.117601,PhysRevB.66.224403}. The illustration of spin-current pumped by the collective spin excitations of magnetic skyrmion is depicted in Fig. \ref{fig:ilus_sp}. By spin-pumping method through the microwave resonance process, the reading-speed racetrack memory has been claimed to be significantly increased as compared to the spin-orbit torque (SOT) or spin-transfer torque (STT) methods \cite{Zhang_2020}. 

Nonetheless, the investigation of the spin-current pumped by the resonated magnetic skyrmion so far is limited only for the application of linearly polarized ac magnetic fields. Meanwhile, an application of circularly polarized ac magnetic fields may have a promising feature since it could modify the magnetization oscillations, which may affect the induced spin-current. Moreover, it was also reported that the uniaxial magnetic anisotropy could control not only the skyrmion phase \cite{Nii2015} but also its dynamics \cite{PhysRevB.94.014406, Chen_2018}. However, to date, the effect of uniaxial magnetic anisotropy on the resonant frequency of the magnetic-skyrmion dynamics and the induced spin currents has not been explored yet.



To investigate the effect of uniaxial anisotropy and polarized ac magnetic fields on the resonant frequency of the magnetic skrymion and the induced spin-current, we have performed micromagnetic simulations on the classical spin systems of the two-dimensional Heisenberg model. Our main results show a profound difference in the spin dynamics of the two types of skyrmion lattice obtained, referred to as SkX type-I and II, emerging at low and large magnetic anisotropy constants, resulting in different natures of the induced spin currents through a resonant process under the polarized microwaves.

\section{Calculation Methods}


In this study, we focus on the classical spin configuration in the two-dimensional square lattice Heisenberg model with $N=128\times 128$. The model is treated classically. We introduce a classical variable $\bm{S}_i$ as an expectation value of the spin at the $i$-th site and the classical Hamiltonian ${\cal H}$, which contains the nearest-neighbor ferromagnetic exchange ($J$), Zeeman coupling ($H_s$), Dzyaloshinskii-Moriya interactions ($D$), and uniaxial anisotropy ($K_z$) terms \cite{Bak_1980,PhysRevB.80.054416,mochizuki,PRADHAN2021167805,Evans_2014,Roldán-Molina_2016},

\begin{eqnarray}
\mathcal{H} = -J \sum_{<i,j>} \bi{S}_i \boldsymbol{\cdot} \bi{S}_j - [\bi{H}_s+\bi{H}'(t)] \boldsymbol{\cdot} \sum_i \bi{S}_i \nonumber \\
- D\sum_i (\bi{S}_i \times\bi{S}_{i+\hat{x}} \boldsymbol{\cdot} \hat{\bi{x}} + \bi{S}_i \times \bi{S}_{i+\hat{y}} \boldsymbol{\cdot} \hat{\bi{y}} ) \nonumber \\
-K_z \sum_i (\bi{S}_i{\cdot} \hat{\bi{z}})^2,    
\end{eqnarray}

\noindent where the static magnetic field is applied normal to the plane, $\bi{H}_s =(0,0,H_z)$, and time-dependent magnetic field $\bi{H}'(t)$ is taken into account to study the spin dynamics. The sum runs over all the lattice points in the real space.

The static spin structures ($H'(t)=0$) at ground state were calculated by minimizing the energy functional, $\varepsilon=\mathcal{H}(\vec{\Omega})$, in which $\vec{\Omega}$ is the classical spin configuration. The energy minimization is obtained by self-consistent iteration. We employed GPU computation with the checkerboard algorithm \cite{PREIS20094468}. The spin over-relaxation method was also used during the self-consistent calculation to accelerate the convergence of the skyrmion phase~\cite{PhysRevD.36.515}. The static spin-correlation $|\bi{S}_q|^2$ is evaluated by Fourier transformation of the real-space spin configuration, $\bi{S}_q=(1/\sqrt{N}) \sum_{j}\bi{S}_{j}e^{-iqr_j}$. 

In the simulations, we set the parameter of Heisenberg exchange coupling $J$ to unity and the magnetic periodicity $\lambda$ as 40 sites which corresponds to the skyrmion size of 20 nm if we consider a typical lattice parameter of 5 \r{A}. The Dzyaloshinskii-Moriya interaction (DMI) is estimated as $D=\sqrt{2}\tan(2\pi/\lambda) \approx 0.22$. The static magnetic field $H_z$ is set to 0.015. All units are in meV, so the applied magnetic field corresponds to $H_z\approx 0.13$ T. Furthermore, the uniaxial anisotropy $K_z$ is varying within the interval of $0-0.25$ meV, with an increment of $K_z$ as $0.005$ meV.

\begin{figure}[b]
\centering
\includegraphics[width=8cm]{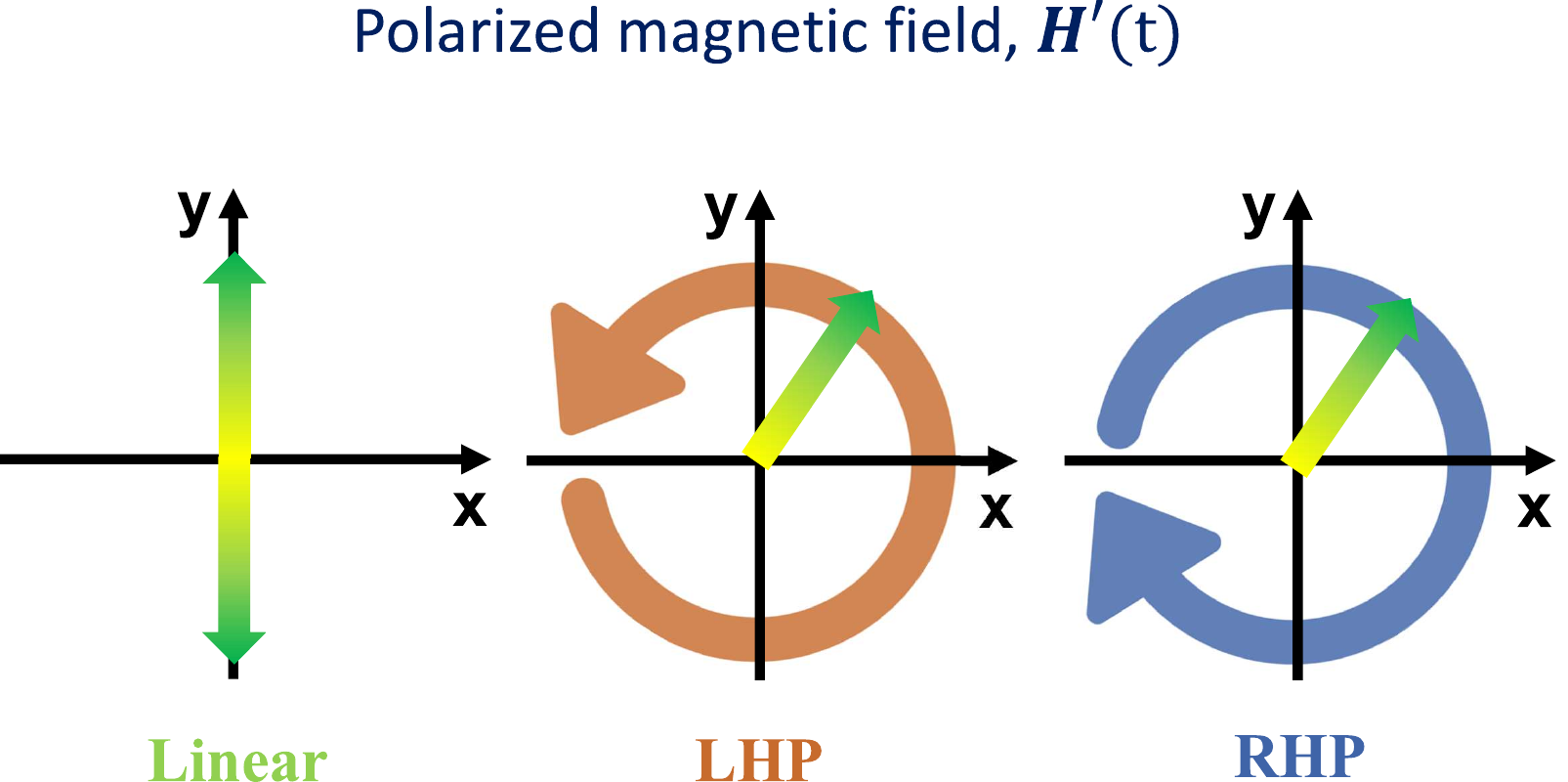}
\caption{Polarized ac magnetic fields $\bm{H}'(t)$ have been applied to excite the magnetic skyrmion during simulations. }
\label{fig:ilus_pf}
\end{figure}

\begin{figure}[b]
\centering
\includegraphics[width=8.5 cm]{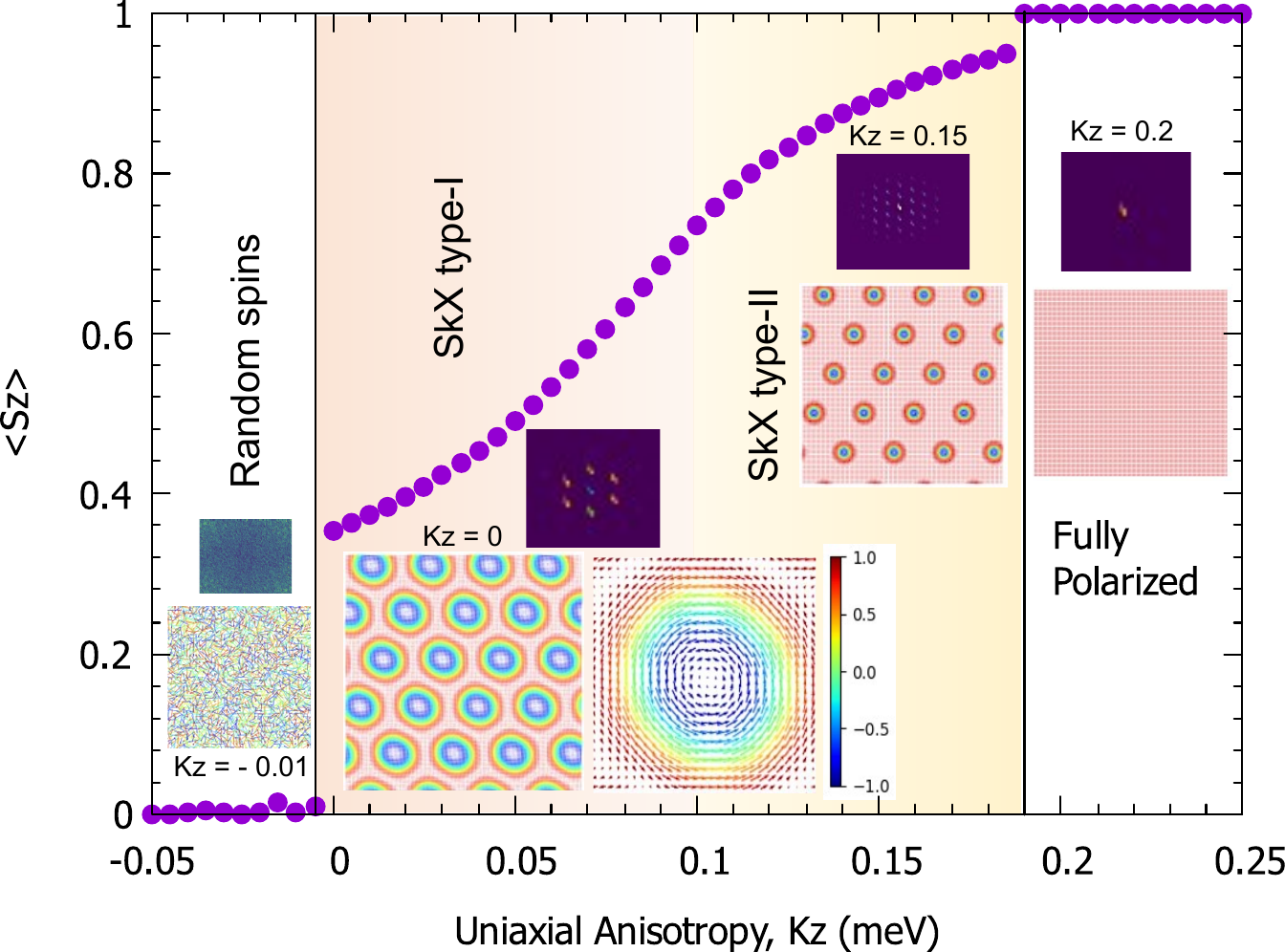}
\caption{A magnetic phase diagram by varying the uniaxial anisotropy. Insets show the spin structure in real space and momentum distribution of spin correlation $|S_q|^2$. Two types of skyrmion lattice of SkX type-I and II with different natures emerge in the presence of uniaxial anisotropy.} 
\label{fig:phasediagram}
\end{figure}

The spin dynamics or collective excitations of the skyrmion lattice are simulated by performing Landau-Lifshitz-Gilbert (LLG) equation as follows~\cite{mochizuki,Evans_2014},

\begin{equation}
\frac{\partial{\bi{S}_i}}{\partial{t}} = -\frac{1}{1+\alpha} \left[\bi{S}_i \times \mathbf{H}^{eff}_i + \frac{\alpha}{S} \bi{S}_i \times (\bi{S}_i \times \mathbf{H}^{eff}_i) \right],
\end{equation}

\noindent where the local spin $\bi{S}_i$ undergoes precession around the effective field $\mathbf{H}^{eff}_i=-\frac{\partial \mathcal{H}}{\partial \bi{S}_i}$, while the Gibert-damping parameter $\alpha$ leads to the relaxation of the spins to the equilibrium state. We fix $\alpha=0.01$ for all simulations. The spin dynamics of $\bi{S}_i(t)$ can be solved numerically using Heun method \cite{PhysRevB.58.14937,Evans_2014}. The absorption spectrum or imaginary part of the dynamical susceptibility, Im$\chi (\omega)$, is calculated from the Fourier transformation of magnetization $\bi{m}(t)=(1/N)\sum_i \bi{S}_i(t)$. The resonant frequency for each excitation modes of skyrmion lattice is obtained by applying the pulsed magnetic fields in-plane ($\bm{H}'(t)||y$) and out-of-plane ($\bm{H}'(t)||z$) \cite{mochizuki}. The time interval of $0-12.4$ ns is selected for all spin dynamics simulations.

The generated spin current density for LLG equation is evaluated according to the equations \cite{RevModPhys.87.1213}, 

\begin{equation}
    J_{sp}\bm{\sigma}(t)=\frac{\hbar A_r}{4\pi} \left( \mathbf{\hat{S}} \times \frac{\partial \mathbf{\hat{S}}}{\partial t}\right)
\end{equation}

\noindent where $\bm{\sigma}$ is the unit vector of the spin-current polarization with its magnitude $J_{sp}$ and $A_r$ is the real spin-mixing conductance of the particular sample, $A_r \approx \mathrm{Re}[g^{\uparrow \downarrow}]$. Then, we evaluate the $z$-polarization of the spin current density $J_{z}\sigma_z(t)$ at one edge of the square lattice of $(x,0)$.  

\begin{equation}
    J_{z}\sigma_z(t)=\frac{1}{l}\int_0^l J_{sp} \sigma_z(t)dx,
\end{equation}

\noindent with $l$ being the length of the edge. Note that the generated spin current is a mixture of AC and DC components. We approximate the amplitude of the AC component by,
\begin{equation}
 \mathrm{Amplitude}(J_z) = \frac{\mathrm{max}(J_z)-\mathrm{min}(J_z)}{2},   
\end{equation}

\noindent the amplitudes are determined when the spin-current oscillations reach convergences ($\approx 8-12.4$ ns). While, the DC component is obtained as follows,
\begin{equation}
    J_{dc}=\frac{1}{T}\int_0^T J_z \sigma_z(t)dt.
\end{equation}

\noindent We evaluate the generated spin current under the linearly ($\eta=0$), left-handed ($\eta=+1$) and right-handed ($\eta=-1$) circularly polarized microwaves [see Fig.~\ref{fig:ilus_pf}], $\bm{H}'(t)=0.01H_z \left[\eta \cos{(\omega_R t)}, \sin{(\omega_R t)},0 \right] $. In the following, we refer the left- and right-handed circularly polarized in-plane ac magnetic fields as LHP and RHP.

\begin{figure*}[t]
\centering
\includegraphics[width=16.5cm]{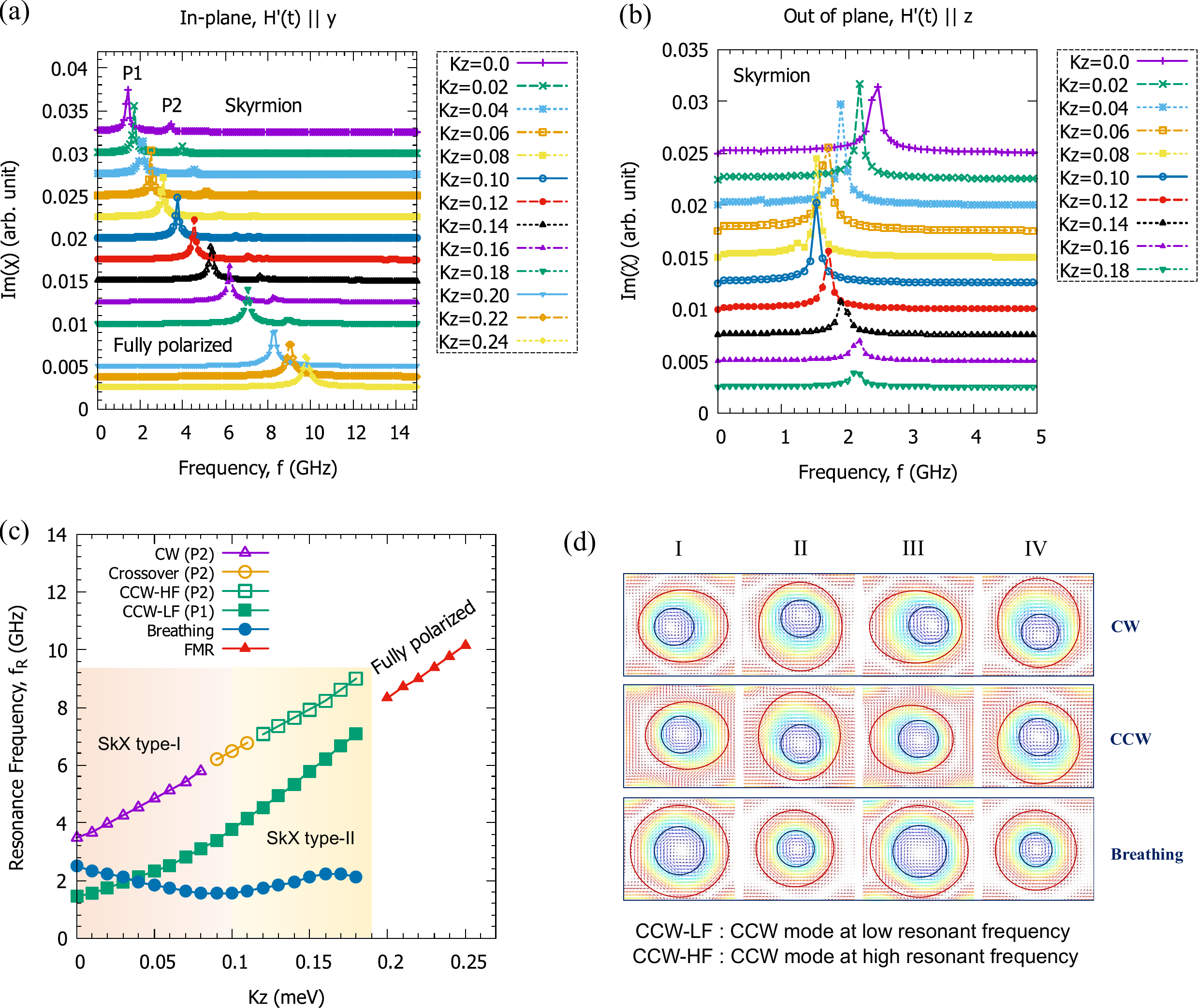}
\caption{Absorption spectrum as a function of frequency at various uniaxial anisotropy constant $K_z$ with pulsed delta function in-plane (a) and out of plane (b). The estimated resonant frequency derived from (a) and (b) is plotted in (c). The spin dynamics of clockwise (CW), counterclockwise (CCW), and breathing modes are illustrated in (d).}
\label{fig:sqw}
\end{figure*}

\begin{figure*}[t]
\centering
\includegraphics[width=16cm]{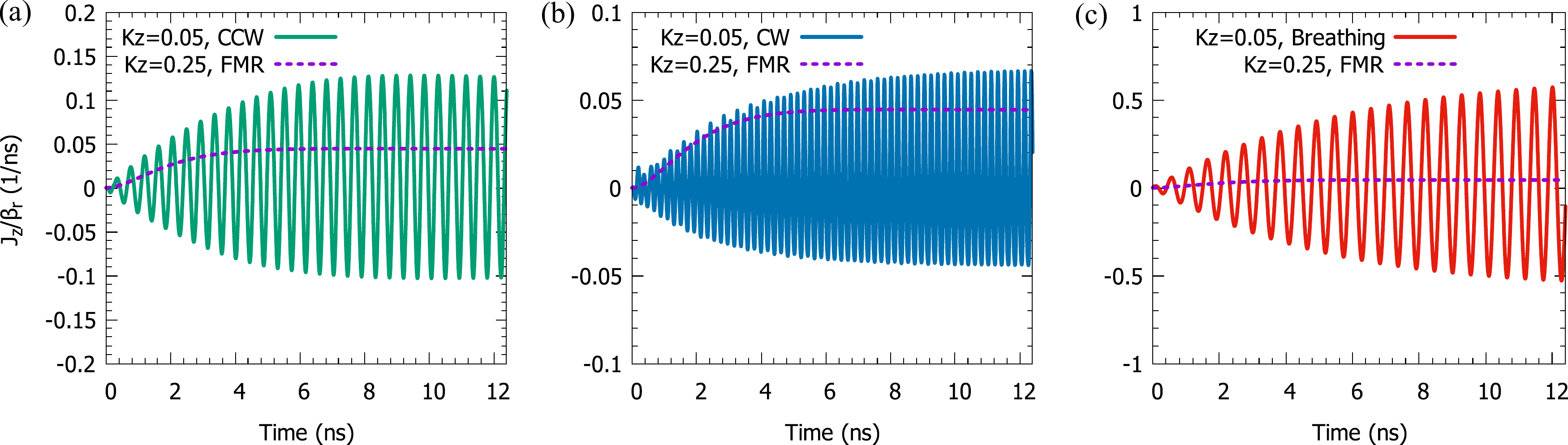}
\caption{The time dependence of spin-currents pumped by resonance process of the skyrmion lattice in (a) CCW-LF, (b) CW, and (c) breathing modes at selected anisotropy constant ($K_z=0.05$ meV) under linearly polarized microwave. Ferromagnetic resonance (FMR) at $K_z=0.25$ meV is also included in each graph as a reference.}
\label{fig:ac_current}
\end{figure*}

\begin{figure*}[t]
\centering
\includegraphics[width=13cm]{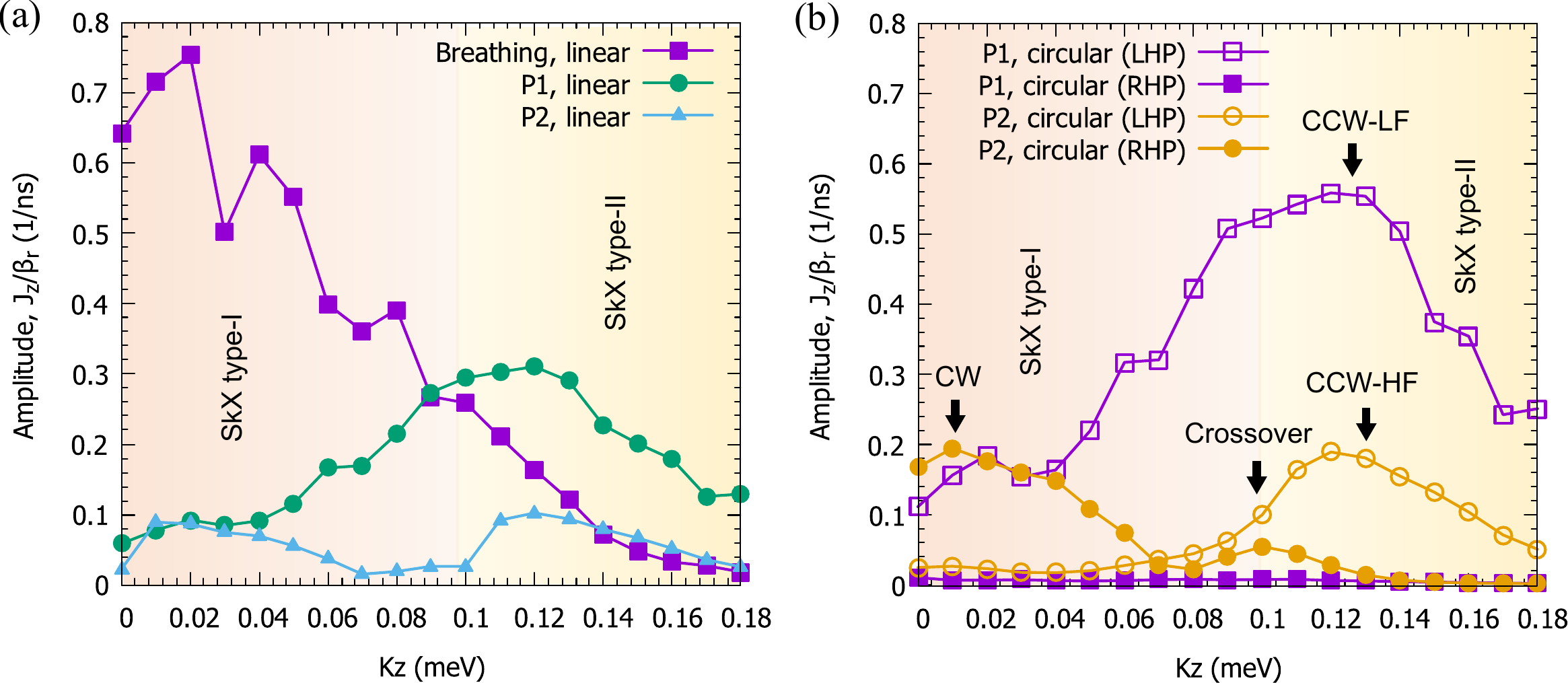}
\caption{Amplitudes of AC spin-current as a function of uniaxial anisotropy $K_z$ under linearly (a) and circularly (b) polarized ac magnetic fields, derived from the time dependence of spin-currents. Determination of amplitudes is approximated by, Amplitude~=~$\frac{{\rm max}(J_z)-{\rm min}(J_z)}{2}$, within the time interval of $8-12.4$ ns where the amplitudes of spin-current oscillations converge.}
\label{fig:amplitudo_jz}
\end{figure*}


\begin{figure*}[t]
\centering
\includegraphics[width=12cm]{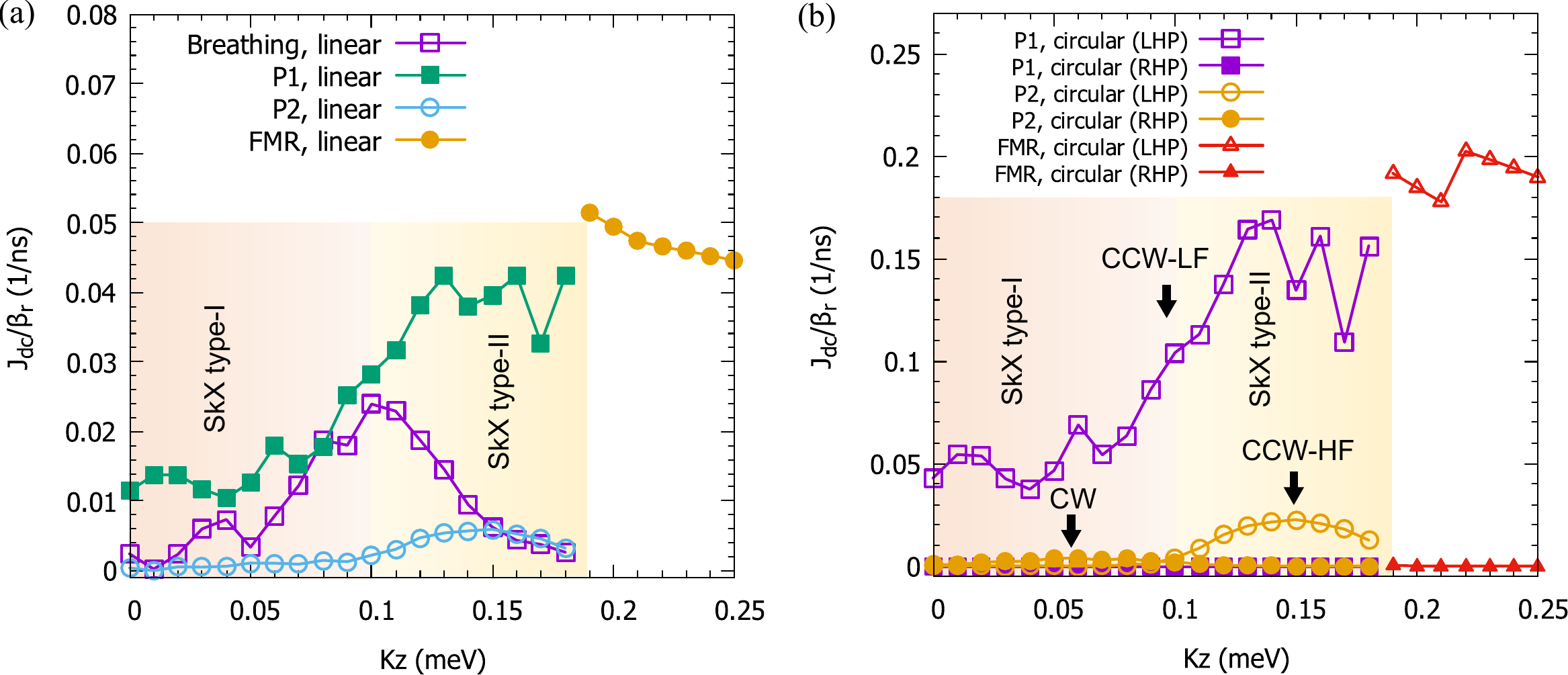}
\caption{Generated DC spin-current under linearly (a) and circularly (b) polarized ac magnetic fields.}
\label{fig:jdc}
\end{figure*}

\begin{figure*}[t]
\centering
\includegraphics[width=12cm]{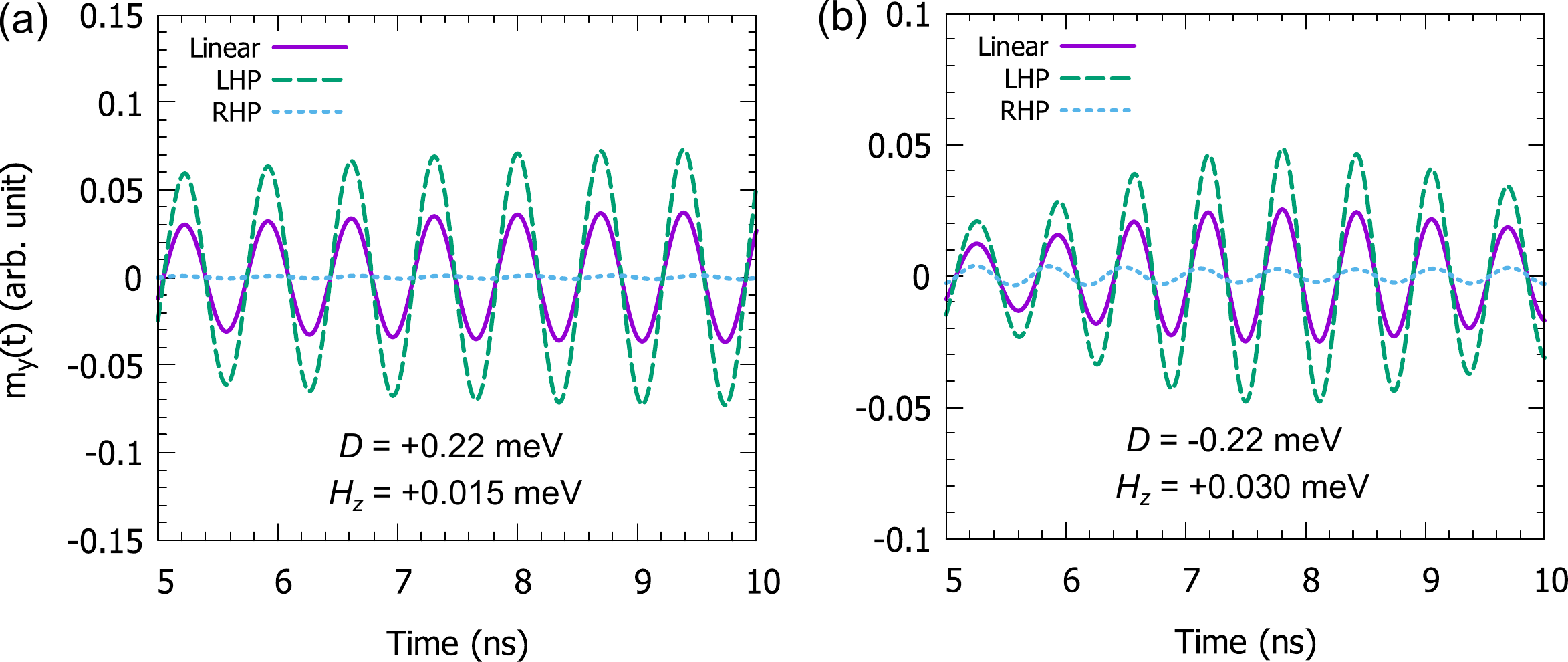}
\caption{Calculated time evolutions of $y$-component of magnetizations in the skyrmion lattice phase, $m_y(t)=(1/N)\sum_i S_{yi} (t)$, under linearly, left-handed (LHP), and right-handed (RHP) circularly polarized in-plane ac magnetic fields at low-frequency CCW modes (CCW-LF) with (a) positive and (b) negative signs of DMI, $D$. Uniaxial anisotropy is absent, $K_z=0$. }
\label{fig:myt}
\end{figure*}

\section{Results and Discussions}

\subsection{Magnetic Phase Diagram}
First, we examine the spin structure at ground state by applying a fixed static magnetic field normal to the plane ($\bm{H}_s \perp x,y$) and varying the uniaxial anisotropy constant, $K_z$. As shown in Fig.~\ref{fig:phasediagram}, in the absence of uniaxial anisotropy, a stable magnetic phase is skyrmion lattice (SkX), indicated by the six-fold pattern in the spin correlation $|S_q|^2$ and also the spin structure in real space (inset figure). By increasing anisotropy, we find a continuous magnetic transition (second-order) with different types of skyrmion lattice, which we refer to as SkX type-I and SkX type-II, followed by a reduction of skyrmion size. The uniaxial anisotropy, by nature, tends to drive the spin along $z$-direction as an easy axis. Consequently, the spins which are polarized parallel to the uniaxial axis become predominant and reduce the skyrmion size~\cite{PRADHAN2021167805,10.1063/5.0181599}. The SkX type-I and II, which will be identified in the later section, have different natures of spin-excitations and the induced spin-current. 

At large anisotropy, there is a first-order magnetic transition from the skyrmion lattice phase to the fully polarized phase, at $K_z\approx 0.19$~meV. This typical first-order transition often occurs in the system with a lower symmetry owing to uniaxial anisotropy~\cite{ASTI198029,PhysRevB.32.264}. In case of the opposite sign of the $K_z$ constant or easy plane, we find a discontinuity around $K_z\approx -0.005$ meV, the spin structure tends to be randomly oriented. This case is out of our interest.    

\subsection{Absorption Spectrum and Resonant Frequency}

Next, we investigate the collective excitations of the skyrmion lattice phase. Shown in Figs. \ref{fig:sqw}(a)-(b) are the absorption spectrum (Im$\chi$) as a function of frequency at various uniaxial anisotropy constant $K_z$. The figures indicate that the in-plane excitations have two absorption peaks indexed by P1 and P2 corresponding to the low and high-lying excitation modes, while the out-of-plane excitations have only a single peak. The obtained resonant frequency of skyrmion lattice phase is within a microwave frequency range of $1-10$ GHz, therefore those excitations are visible to be probed by the microwave spectroscopy~\cite{onose_lett,Satywali2021-mg,Schwarze2015,Okamura2013,PhysRevB.94.014406}.


We plot the estimated resonant frequencies as a function of uniaxial anisotropy in Fig.~\ref{fig:sqw}(c). From the nature of spin-current under the polarized microwave, which will be described in the next section, we find that the low-lying resonant frequency of in-plane excitations (P1) belongs to the CCW rotation in a wide range of anisotropy constants, called CCW-LF. On the other hand, the high-lying resonant frequency of in-plane excitations (P2) has different modes depending on the anisotropy strength, the CW rotation at low anisotropy ($K_z<0.1$ meV - SkX type I) and CCW rotation at large anisotropy ($K_z>0.1$ meV - SkX type II). In addition, the CW and CCW rotations coexist (crossover) at the boundary of two skyrmions. In the following, we define CCW-LF and CCW-HF as CCW excitations emerging at low (P1) and high (P2) resonant frequencies in the absorption spectrum. The typical spin dynamics of clockwise (CW), counterclockwise (CCW), and breathing modes at low anisotropy constants are illustrated in Fig.~\ref{fig:sqw}(d) 

In addition, we find that the resonant frequency of the in-plane excitation modes is shifted to the higher frequency monotonously (blueshift) as the anisotropy constant increases. A monotonically increasing resonant frequency of in-plane excitation modes indicates that the anisotropy field enhances the skyrmion core rotation since the dynamics are dominated by the spins, which are perpendicular to the anisotropy field, resulting in a maximized torque.


In the out-of-plane excitations, we find different natures of the breathing modes for SkX type-I and II. As shown in Fig.~\ref{fig:sqw}(c), in the absence of anisotropy, the resonant frequency of breathing mode is 2.6 GHz, lying in between the frequencies of the two in-plane rotation modes, CCW-LF (1.5 GHz) and CW (3.5 GHz) excitations. The hierarchy of these excitation modes is consistent with the universal behavior of Bloch-type skyrmion excitations~\cite{Schwarze2015}. However, at anisotropy constant $K_z>0.04$ meV, we find that the resonant frequency of the breathing mode becomes the lowest compared to those rotation modes, suggesting that the uniaxial anisotropy plays an essential role in interchanging the hierarchy of the excitations. This kind of interchange was also observed in the Neel-type skyrmion lattice of GaV$_4$S$_8$ owing to the uniaxial anisotropy \cite{PhysRevB.94.014406}. Furthermore, we find that the resonant frequency of the breathing mode is shifted to the lower frequency at $K_z<0.1$~meV (redshift) in the SkX type I, and then raised to the higher frequency at $K_z>0.1$~meV (blueshift) in the SkX type II. 




As a comparison, we also estimate the resonant frequency in the fully polarized phase region where all spins are parallel to the static field direction, $\bm{H}_s || z$. In this case, the oscillation of the magnetic field is applied perpendicular to the polarized spins, $\bm{H}'(t) || y$. We find that the ferromagnetic resonance (FMR) shows a simple linear behavior against uniaxial anisotropy, which is a typical feature of FMR under such an anisotropy~\cite{Kachkachi2007}. Generally, in the present simulation, the resonant frequency of in-plane excitation modes of both types of skyrmion lattices under uniaxial anisotropy is increasing, but is still lower than that of FMR. Except near the transition $K_z\approx0.19$ meV, where the resonant frequency of CCW-HF mode exceeds the FMR. However, note that the intensity of CCW-HF mode is weaker than that of CCW-LF mode, even with FMR, as shown in Fig. \ref{fig:sqw}(a).

\subsection{Evaluation of Spin-Current}
Finally, we evaluate the spin-current with $z$-spin polarization pumped by the skyrmion lattice as illustrated in Fig. \ref{fig:ilus_sp}. Here, the evaluation of the generated spin-current is carried out only at one edge, $(x,0)$, which is adequate to capture the essence, for our concern. Our calculation shows that the generated spin-current is a mixture of AC and DC components arising from the spin excitations of the skyrmion lattice (see Fig. \ref{fig:ac_current}). On the other hand, the FMR only produces the DC component. These features also have been pointed out in Ref.~\cite{Zhang_2020}.  

The obtained spin currents in the present simulations have been normalized by the constant related to the real spin-mixing conductance of non-magnetic metals,

\begin{equation}
    \beta_r = \frac{\hbar A_r}{4\pi}.
\end{equation}

\noindent Note that spin-mixing conductance is a complex value. However, theoretical \f{ab initio} calculations and phase randomization at the scattering interface indicate that only the real part of spin-mixing conductance dominates the physics in the order of $A_r\approx \mathrm{Re}[g^{\uparrow \downarrow}]= k_\mathrm{F}^2/4\pi$ \cite{RevModPhys.87.1213,PhysRevB.66.224403,PhysRevB.65.220401, PhysRevB.71.064420, PhysRevB.76.104409}. Suppose we consider a heavy metal of gold (Au) as a non-magnetic layer being a perfect sink, $A_r=1.147 \times 10^{19}$ m$^{-2}$~\cite{Zhang_2020}. Thus, the estimated spin-currents pumped by a magnetic skrymion in the present simulations are in the order of $10^{-7}$ A/m$^2$.    

To capture the effect of uniaxial anisotropy on the induced spin-current, we plot the amplitude of AC spin-current against the uniaxial anisotropy as shown in Fig~\ref{fig:amplitudo_jz}(a). We first start with an application of a linearly polarized microwave. In the absence of anisotropy, we find that the breathing mode has a considerably huge amplitude in comparison with the in-plane excitation modes (P1 and P2). Nonetheless, as the uniaxial anisotropy increases, the amplitude of the spin-current of the breathing mode shows a decreasing trend, which means that such an anisotropy suppresses the oscillation of the skyrmion core along the uniaxial axis. At $K_z>0.1$~meV (SkX type-II), the amplitude of spin-current in the P1 excitations becomes more profound.   

Next, we examine the spin-current pumped by the collective excitations of skyrmion lattice under the circularly polarized in-plane ac magnetic fields. Please note that an application of circularly in-plane ac magnetic fields excites only CCW and CW modes. As shown in Fig~\ref{fig:amplitudo_jz}(b), the amplitude of spin-current arising from the in-plane P1 excitations under the LHP microwave is strongly intense in the presence of anisotropy, while it is quenched by the RHP microwave, indicating that the P1 excitations correspond to the CCW mode lying at low resonant frequency (CCW-LF). On the other hand, the P2 excitations have different types of spin dynamics depending on the anisotropy strength. The CW mode appears at low anisotropy constant ($K_z<0.08$ meV) indicated by the intensely induced spin-current under RHP microwave. The CCW mode lying at high resonant frequency (CCW-HF) emerges at large anisotropy constant ($K_z>0.12$ meV), triggered by the LHP microwave. The CW and CCW modes coexist at the boundary of SkX type-I and II (crossover).



We also extract the DC component from the oscillating spin-current, as shown in Fig. \ref{fig:jdc}. Under linearly polarized microwave [Fig.~\ref{fig:jdc}(a)], we find the P1 excitations exhibit more intense DC spin-current in comparison with breathing and P2 excitations, in a wide range of anisotropy. The uniaxial magnetic anisotropy enhances the DC spin-current induced by the P1 excitations monotonously. However, the resulting spin-current is still lower than that in the ferromagnetic resonances. It is expected, since in the ferromagnet, all spins are polarized in the same direction (along the uniaxial axis), yielding a maximized spin polarization that pumps the spin-current. On top of that, we find that the DC spin-current induced by breathing mode has a positive trend at a small anisotropy ($K_z< 0.1$ meV), and undergoes a negative trend at a large anisotropy ($K_z> 0.1$ meV) in concomitant with the increase of the induced DC spin-current by the P2 excitations.



Furthermore, we find that the spin currents pumped by CCW-LF and CCW-HF modes are considerably enhanced under LHP microwaves [Fig.~\ref{fig:jdc}(b)]. They are approximately four times stronger than the DC spin currents produced under a linearly polarized microwave. In contrast, the pumped spin currents become very weak under RHP microwaves. This also occurs in the ferromagnetic resonance. Strong enhancement of the obtained spin-currents may be originating from the magnetization oscillations, which are significantly enhanced under the LHP microwave but quenched under the RHP microwave as shown in Fig.~\ref{fig:myt}(a), regardless of the sign of Dzyaloshinskii-Moriya interaction (DMI), see Fig.~\ref{fig:myt}(b), which implies that the LHP microwave enhances the magnetization oscillations for both left- and right-handed chirality of the skyrmion crystals. Moreover, we do not observe the DC spin-current induced by the CW mode at low anisotropy. This type of excitation produces only the AC spin-currents as indicated in Fig.~\ref{fig:amplitudo_jz}.   

Our present simulation suggests that there are profound differences in the nature of the skyrmion lattices at small ($K_z<0.1$ meV) and large ($K_z>0.1$ meV) magnetic anisotropy constants. The induced spin currents under polarized microwaves could distinguish the nature of their spin excitations. Furthermore, circularly polarized microwaves could also be regarded as a switch of spin-current generation, either induced by CW/CCW-LF modes at a low anisotropy or CCW-HF/CCW-LF at a large anisotropy constant. The anisotropy strength can be tuned by rotating the crystallography axes of the sample or by applying mechanical stress to the sample. In addition, the circularly polarized microwaves also enhance the amplitudes of alternating and direct spin currents. These features may have a great advantage for skyrmion-based spintronic applications in the future.

\section{Summary}
In summary, we have performed theoretical simulations on the spin current generated by the spin-pumping method through the resonant process in a magnetic skyrmion with a lack of inversion symmetry. The effects of uniaxial magnetic anisotropy and under polarized magnetic fields have been examined. We obtain two skyrmion lattices, namely SkX type I and II, emerging at low ($K_z<0.1$ meV) and large ($K_z>0.1$ meV) magnetic anisotropy constants. The SkX type-I exhibits a CW/Breathing/CCW-LF mode at a very low anisotropy, and the hierarchy of these excitations interchanges to CW/CCW-LF/breathing mode at $K_z\sim 0.04$~meV. Meanwhile, the SkX type-II shows different kinds of spin excitation in which the clockwise mode is obliterated, while counterclockwise modes emerge at both low and high resonant frequencies. This suggests that the magnetic anisotropy plays an essential role in spin dynamics. The anisotropy strength can be tuned in the experiment, either by rotating the crystallography axes of the sample or by mechanical stress. Furthermore, the nature of those spin excitations can be clearly distinguished by the left- and right-handed circularly polarized microwaves (LHP and RHP). Such polarized microwaves can act as a switch for the spin current generation. In addition, an application of circularly polarized microwaves could enhance the amplitude of alternating and direct spin currents induced. These features may have a great application for the spintronics device.



\begin{center}
\bo{ACKNOWLEDGMENTS}
\end{center}

This theoretical study was supported by PUTI Q1 Research Grant No. NKB-481/UN2.RST/HKP.05.00/2023, from Universitas Indonesia, in the period of 2023/2024. 

\bibliographystyle{apsrev4-2}
\bibliography{reference_puti}

\end{document}